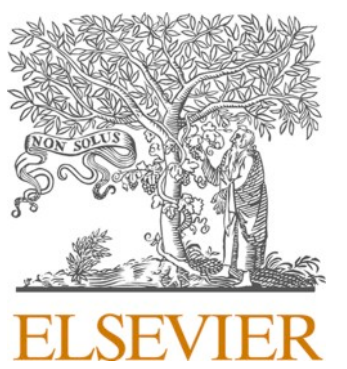

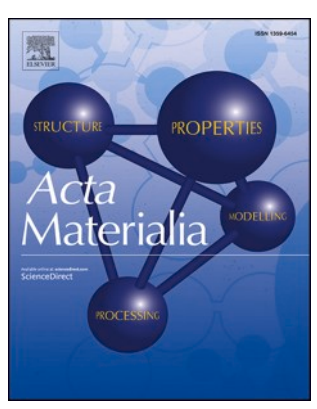

Full length article

# Multimodal deep learning framework to predict strain localization of Mg/LPSO two-phase alloys

Daiki Kuriki, Fabien Briffod, Takayuki Shiraiwa *, Manabu Enoki

*Department of Materials Engineering, School of Engineering, The University of Tokyo, 7-3-1 Hongo, Bunkyo-ku, Tokyo 113-8656, Japan*

A B S T R A C T

This study proposes a method for predicting three-dimensional (3D) local strain distribution under compressive deformation of as-cast Mg/LPSO two-phase alloys from 3D microstructure images. The 3D local strain distribution was obtained by applying the digital volume correlation method to X-ray CT images before and after compression tests. Three microstructure descriptors were extracted from the 3D microstructure images around each strain measurement point: volume fractions of the phases, persistent diagrams that can express the connectivity of the phases, and two-phase spatial correlation that can express the spatial distribution of the phases. A deep learning model was then constructed to predict local strain from the three microstructure descriptors. Since two types of descriptors were used in this study, numerical data and image data, multimodal deep learning was employed to make predictions. Thus, the use of multiple microstructure descriptors enabled predictions to be made with higher accuracy than when predictions were made from a single descriptor. Feature importance of the descriptors was assessed through correlation analysis and occlusion sensitivity analysis. The results revealed that high strain tended to occur in the region where the hard phase, LPSO phase, had a large elongated phase oriented at a 45° direction to the loading direction. This result is consistent with other previous studies and indicates that the proposed method is effective in elucidating the relationship between the microstructure and the deformation behavior of the material.

## 1. Introduction

Since magnesium (Mg) and its alloys are the lightest of all structural metals and have excellent specific strength, they are expected to be applied as structural materials for transportation equipment to improve energy efficiency. However, conventional Mg alloys have disadvantages such as low strength and a tendency to ignite, particularly in comparison to steels and aluminum alloys currently used as structural materials for transportation equipment. Under such circumstances, Mg alloys with long-period stacking ordered (LPSO) phase developed by Kawamura et al. [1] have attracted considerable attention. These alloys show remarkably high strength among Mg alloys because of its unique deformation mechanism. Generally, when Mg alloys are compressed under loading conditions that prevent basal slip, twinning deformation occurs, but the LPSO phase undergoes a unique deformation called kink deformation instead of twinning [2]. This kink deformation is an obstacle to basal slip, resulting in high strength. It is important to clarify the details of the deformation mechanism of the material for its practical use as a structural material. In these materials, many studies have observed the deformation behavior on the nano-to-micro scale using a transmission electron microscopy (TEM). However, for application as a structural material, it is also important to understand the deformation behavior across the micro-to-mesoscopic scale, but such studies are still scarce. Here, microscale refers to the level of individual crystalline grains, focusing on their shape, size, and orientation. This scale typically ranges from sub-micrometer to a few micrometers. Mesoscale denotes larger features such as phase boundaries and the arrangement or alignment of multiple grains, which can span several grains and typically have dimensions on the order of hundreds of micrometers. From this perspective, high-resolution digital image correlation (HR-DIC) have recently been performed to reveal the deformation behavior at the mesoscale, which corresponds to the scale of phase distribution [3]. One notable outcome of this approach has been the demonstration that the hard LPSO phase preferentially undergoes plastic deformation in as-cast Mg/LPSO two-phase alloys. Furthermore, Harjo et al. [4] used in-situ neutron diffraction to observe the deformation behavior of a hot-extruded Mg/LPSO two-phase alloy during tensile test and found that the strength of both the α-Mg and LPSO phases increased through different microstructure development. Thus, a better understanding of the deformation behavior at each phase level is important because it will lead to the realization of high reliability, which is essential for structural materials, and also to the creation of better materials through optimization of manufacturing conditions.

* Corresponding author.
*E-mail address:* shiraiwa@rme.mm.t.u-tokyo.ac.jp (T. Shiraiwa).

Most conventional studies on the deformation of metallic materials have been conducted by surface observation in two dimensions (2D). Recently, three-dimensional (3D) observation using X-ray computed tomography (CT) has been actively conducted, providing valuable insights. Toda et al. [5] utilized synchrotron radiation CT to examine voids generated in dual-phase (DP) steel during the tensile test. 3D internal observations revealed that the main mechanism is the rapid growth of voids generated by martensitic cracking. Such 3D observations are particularly useful for materials with complex microstructures, such as two-phase microstructures. Moreover, beyond qualitative 3D observation, several methods for quantitatively evaluating 3D deformation behavior have emerged. One such method is digital volume correlation (DVC), an extension of the DIC method to 3D. It allows the measurement of 3D displacement and strain distribution within materials using images obtained through X-ray CT or other 3D imaging techniques. In metallic materials, this method has been successfully applied to aluminum (Al) alloys [6,7], characterized by their status as light metals that are easily penetrated by X-rays. Currently, there are very few examples of DVC being applied to Mg alloys, but since Mg is a light metal like Al, it is considered to have high applicability.

To understand deformation behavior, it is important not only to measure strain distributions with DVC but also to relate them to the microstructure. In recent years, extensive research has focused on predicting material properties and mechanical behavior from microstructures using machine learning [8-10]. To achieve a more accurate prediction of material properties and mechanical behavior, the improvement of microstructure quantification methods and machine learning models are key elements. The descriptors used to quantify microstructures are categorized into two main types: numerical data, such as grain size [9] and grain boundary characteristics [8], and image data, such as two-point spatial correlation [10] and persistent homology in 2D or 3D. Machine learning models utilizing numerical data inputs offer the advantage of easier physical interpretation of the results. The two-point spatial correlation, which describes the relative positioning of grains or phases, has been effectively used in material discovery by accurately regressing structure-property relationships [10]. Persistent homology analysis, a method within topological data analysis, has gained attention for microstructural quantification. This technique uses persistence diagrams to evaluate the shape of the data and finds applications in materials science, such as assessing phase connectivity in microstructural images [11,12].

In addition, multimodal deep learning models that can simultaneously process different types of inputs, such as numerical and image data, are also recognized as advanced models. Due to the advancement of observation, measurement devices, and analytical techniques, there is an increasing amount of information obtained from materials. Therefore, it is important to integrate the obtained information regardless of its format and use it for prediction. In the field of materials science, although studies utilizing multimodal deep learning are currently not very common, there are examples such as studies predicting the mechanical properties of composite materials from microstructure images by a scanning electron microscope (SEM) and numerical data on composition [13]. Additionally, there is research focused on optimizing the steel-making casting process, where a model was constructed to predict the cooling water temperature from the temperature distribution of the slab (image data) and process conditions (numerical data) [14]. These studies have reported that the prediction accuracy significantly varies with different methods of integrating data [13], and improvements in computational speed were observed with varying network structures [14]. It is believed that enhancing the efficiency of network structures and learning methods can further advance the potential of multimodal deep learning.

In this study, we proposed a deep learning model to predict local strain from microstructures in order to elucidate mesoscale deformation behavior, which is an important aspect to ensure high reliability for practical use of Mg/LPSO two-phase alloys as structural materials. For materials with complex microstructures such as Mg/LPSO two-phase alloys, we considered that observation and analysis in 3D would be useful, so we measured 3D strain distribution using a combination of X-ray CT and DVC, a method with limited research in Mg alloys. From the 3D microstructure images, three microstructure descriptors – volume fractions, persistence diagrams, and two-point spatial correlation – were extracted, and a deep learning model was constructed to output local strain. Since two types of descriptors were used, numerical data and image data, a multimodal deep learning model was employed. To effectively utilize the three microstructure descriptors, a learning method utilizing fine-tuning was proposed. This method achieved significantly higher accuracy in predicting local strain compared to predictions from a single microstructure descriptor. Furthermore, through feature importance analysis, insights into the relationship between microstructure and local strain were gained. The proposed approach can be easily extended and applied in studies aiming to predict material properties and mechanical behavior from various information about materials, thereby elucidating the underlying relationships.

## 2. Materials and methods

### 2.1. Materials and experimental methods

The material used in this study was an as-cast Mg/LPSO two-phase alloy $Mg_{94}Zn_2Y_4$. Table 1 shows the chemical composition. The material was estimated to contain about 40% LPSO phase. Compression specimens of 3 × 3 × 5 (L×W×H) mm were cut from the cast alloy. A compression test was conducted using a Deben MT5000 tensile/compression stage at room temperature atmosphere, employing a crosshead speed of 0.5 mm/min. The loading direction was parallel to the longitudinal direction of the specimen. The test concluded when the compressive stress reached 300 MPa.

The specimen was observed before and after the compression test using a Zeiss Xradia Versa 520 X-ray microscope (XRM). There is generally a trade-off between the area scanned by the XRM and the spatial resolution. In this study, it was necessary to reduce the voxel size to observe enough material microstructure patterns to perform DVC. Therefore, the XRM imaging area was limited to a portion of the lower part of the specimen where significant deformation was observed in the compression test. The voxel size was approximately 0.68 μm. The voltage and power of the X-ray source were set to 80 kV and 7 W, respectively. The specimen was rotated 360 degrees, during which 2401 projection images were acquired. The exposure time per image was 27 seconds.

### 2.2. Digital volume correlation

DVC was performed using 3D reconstructed images before and after the compression test. To save computational resources, images with a size of 600 × 600 × 400 (L×W×H) voxels were cut out from the reconstructed images before and after the test, respectively. Note that this image size corresponds to about 408 × 408 × 272 μm. When cropping the images, care was taken to extract as much of the same area as possible before and after compression. The noise in the images was reduced by a Gaussian filter (filter size 5 × 5, standard deviation 1.6) to improve the accuracy of DVC. The Matlab-based program ALDVC [15] (version 1.0.4) was used to perform DVC. This software uses a hybrid algorithm that combines the advantages of the local method with its superior calculation speed and the global method with its superior calculation accuracy. When performing DVC measurements, the subset was set as a cube with 80 voxels per side, and the step size, or distance between measurement points, was 20 voxels. The output of ALDVC was the strain values of six independent components. In this study, the equivalent strain ($\varepsilon_{eq}$) was calculated from them at each measurement point using the following equation.

**Table 1**
Chemical composition of Mg/LPSO two-phase alloy (in mass%).

| Mg | Zn | Y |
|---|---|---|
| Bal. | 2.60 | 4.68 |

$$\varepsilon_{eq} = \frac{\sqrt{2}}{3}\left[\left(\varepsilon_{xx}-\varepsilon_{yy}\right)^2+\left(\varepsilon_{yy}-\varepsilon_{zz}\right)^2+\left(\varepsilon_{zz}-\varepsilon_{xx}\right)^2+6\left(\varepsilon_{xy}^2+\varepsilon_{yz}^2+\varepsilon_{zx}^2\right)\right]^{\frac{1}{2}} \tag{1}$$

### 2.3. Microstructure descriptors

To clarify the relationship between microstructure and local strain, a cube image of 101 voxels per side centered at the measurement point was extracted at each strain measurement point (hereinafter called RVE), and the local strain was predicted from the RVE. The method of RVE extraction is shown in Fig. 1: at a measurement point placed every 20 voxels (about 13.6 μm), a microstructure image of 101 voxels (about 68.7 μm) per side was extracted from the surrounding area (Fig. 1 (a)). Each phase can be distinguished by the contrast of the 3D image. The RVE was segmented into each phase (Fig. 1 (b)) using Trainable Weka Segmentation [16], a plug-in for ImageJ [17], for the extraction of microstructure descriptors described below. Trainable Weka Segmentation employs a Random Forest algorithm initialized with 200 trees and two random features per node to classify and segment the image data.

Since predicting local strain directly from RVE is challenging, microstructure descriptors were extracted from within the RVE and utilized for prediction. Three types of descriptors were computed in this study: volume fractions, persistence diagrams, and two-point spatial correlations. A schematic of descriptor extraction in the RVE is provided in Fig. 2.

Volume fractions can be easily calculated from segmented images,

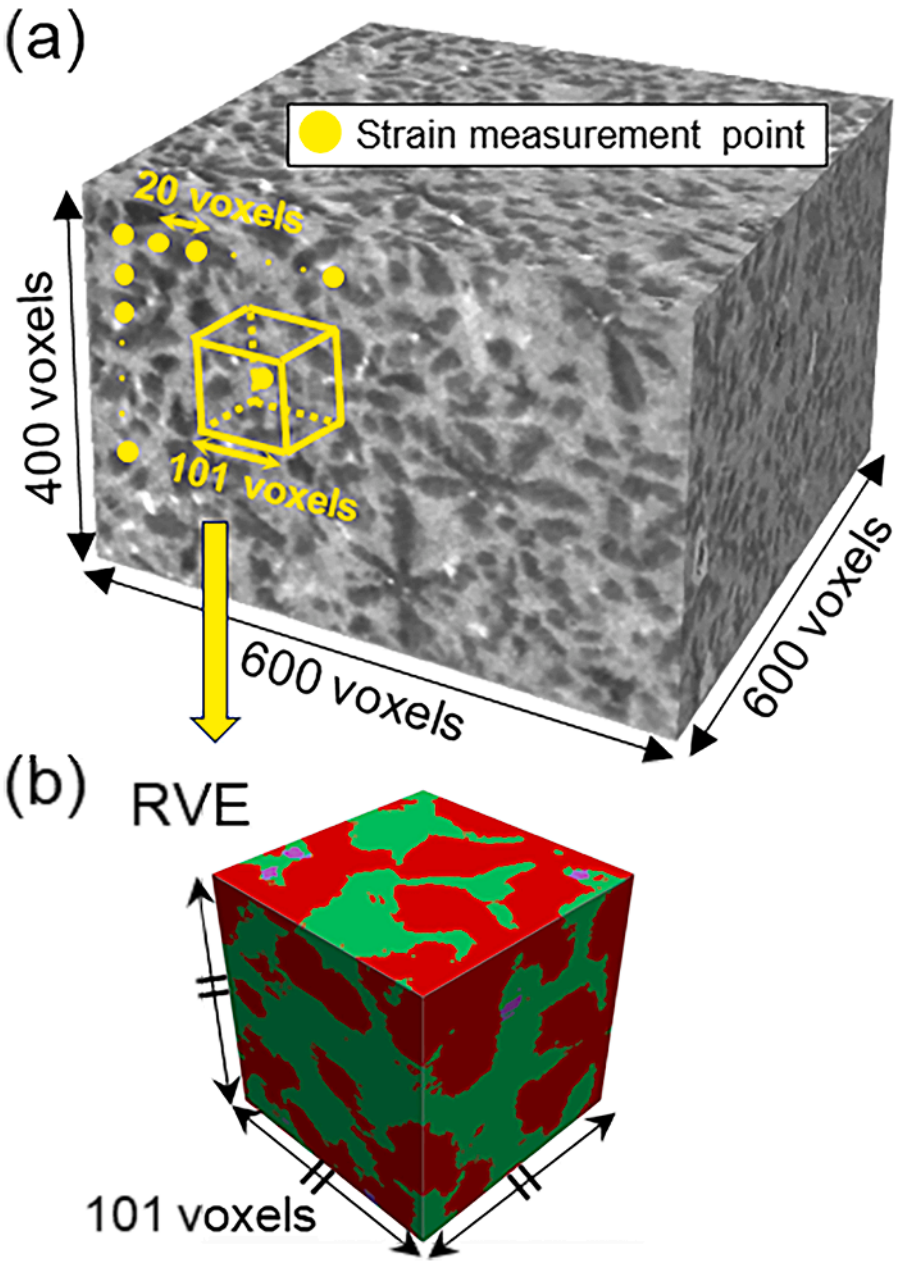

**Fig. 1.** (a) Schematic of spacing of strain measurement points (indicated by yellow dots in the figure) and RVE size in 3D microstructure image used for reference of DVC calculation. (b) Example of RVE extracted from the perimeter of a strain measurement point.

and in this study the volume fractions of the α-Mg phase, LPSO phase, inclusion, and pore in the RVE were calculated (Fig. 2 (b)).

Zero-order persistence diagrams were employed to assess the connectivity of the phases. The persistence diagrams for the α-Mg and LPSO phases were calculated and used as descriptors (Fig. 2 (c)). Python-based software HomCloud [18] (version 4.1.0) was employed for the calculation. When computing the persistence diagram for the α-Mg phase, a two-phase image was binarized with the α-Mg phase as True and other phases as False. Similarly, for the persistence diagram of the LPSO phase, the LPSO phase was binarized to True and other phases to False.

Since persistent homology analysis is a topology-based method, it cannot include information on direction. However, anisotropy is important in the deformation behavior of metallic materials. To address this, we used two-point spatial correlation [19] as a microstructure descriptor. The two-point spatial correlation function averages the correlation between arbitrary points, representing the spatial distribution of phases. In the spatial correlation function, we first consider the probability $m_s^n$ that location $s$ of the object in question is in state $n$. This is called the microstructure function. The microstructure function $m_s^n$ has the following properties:

$$\sum_{n=1}^{N} m_s^n = 1,\ m_s^n \geq 0 \tag{2}$$

where $N$ is the total number of possible states. When evaluating the microstructure function for a material's microstructural image, each pixel $s$ is associated with states $n$ such as α-Mg, LPSO phases, or defects, allowing for the evaluation of phase and defect distributions. The two-point spatial correlation function is defined as:

$$f_t^{nn'} = \frac{1}{S}\sum_{s=0}^{S-1} m_s^n m_{s+t}^{n'} \tag{3}$$

where $t$ is the vector between two points. In other words, the two-point spatial correlation function represents the probability that two points separated by vector $t$ are in states $n$ and $n'$, respectively. The correlation function for $n = n'$ is called autocorrelation, and for $n \neq n'$, it is called cross-correlation. In this study, the autocorrelation of the LPSO phase was calculated based on 3D images. Due to the high computational cost associated with using 3D images as input for deep learning, this study used two cross sections of the zx plane and yz plane as descriptors, These sections include information along the loading direction (z-direction) (Fig. 2 (d)). The Matlab Spatial Correlation Toolbox: Release 3.1 [20] was used to calculate the two-point spatial correlation.

### 2.4. Deep learning model

A deep learning model was developed to predict local strain from microstructure descriptors extracted from the RVE. To handle both numerical data (e.g., volume fraction) and image data (e.g., persistence diagrams and two-point spatial correlations), a multimodal deep learning approach was employed. Fig. 3 illustrates the overall structure of this multimodal model.

In the preprocessing stage, numerical data was standardized, and image data was resized to 32 × 32 pixels and converted to grayscale. The model integrates two types of neural networks: Fully Connected Neural Networks (FCNNs) for numerical data and Convolutional Neural Networks (CNNs) for image data. Specifically, FCNNs were used to process numerical data like volume fractions. CNNs were used for feature extraction from persistence diagrams and two-point spatial correlations. For persistence diagrams, separate CNNs extracted features for α-Mg and LPSO phases. These features were concatenated and fed into a fully connected (FC) layer to form a unified feature vector. Similarly, separate CNNs processed the two-point spatial correlations for the zx and yz planes, and the resulting vectors were combined in an FC layer. The extracted feature vectors, each of size 64, were merged into a single

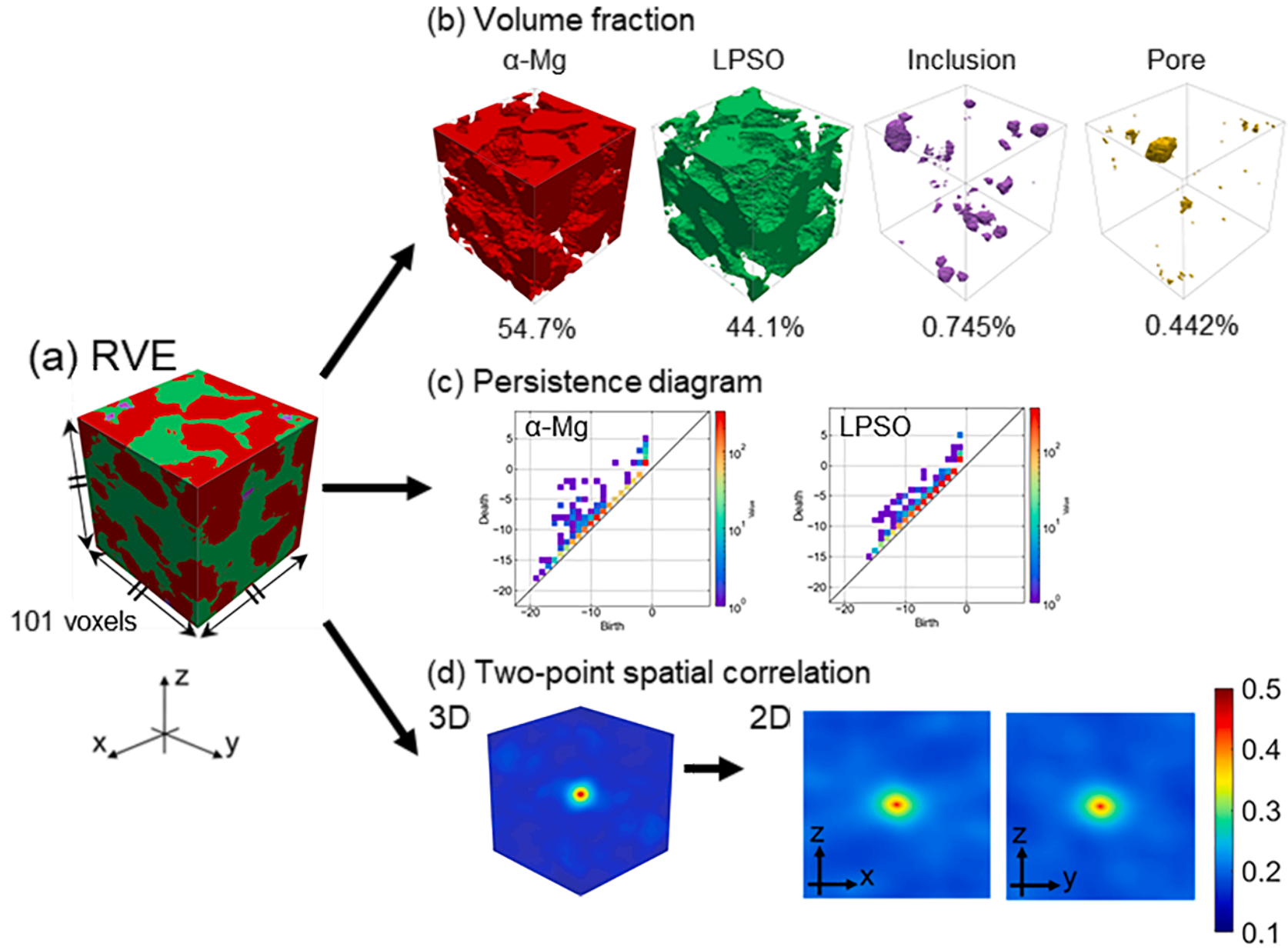

**Fig. 2.** Schematic of microstructure descriptors extraction. (a) RVE example, (b) Volume fraction (c) Persistence diagram (d) Two-point spatial correlation.

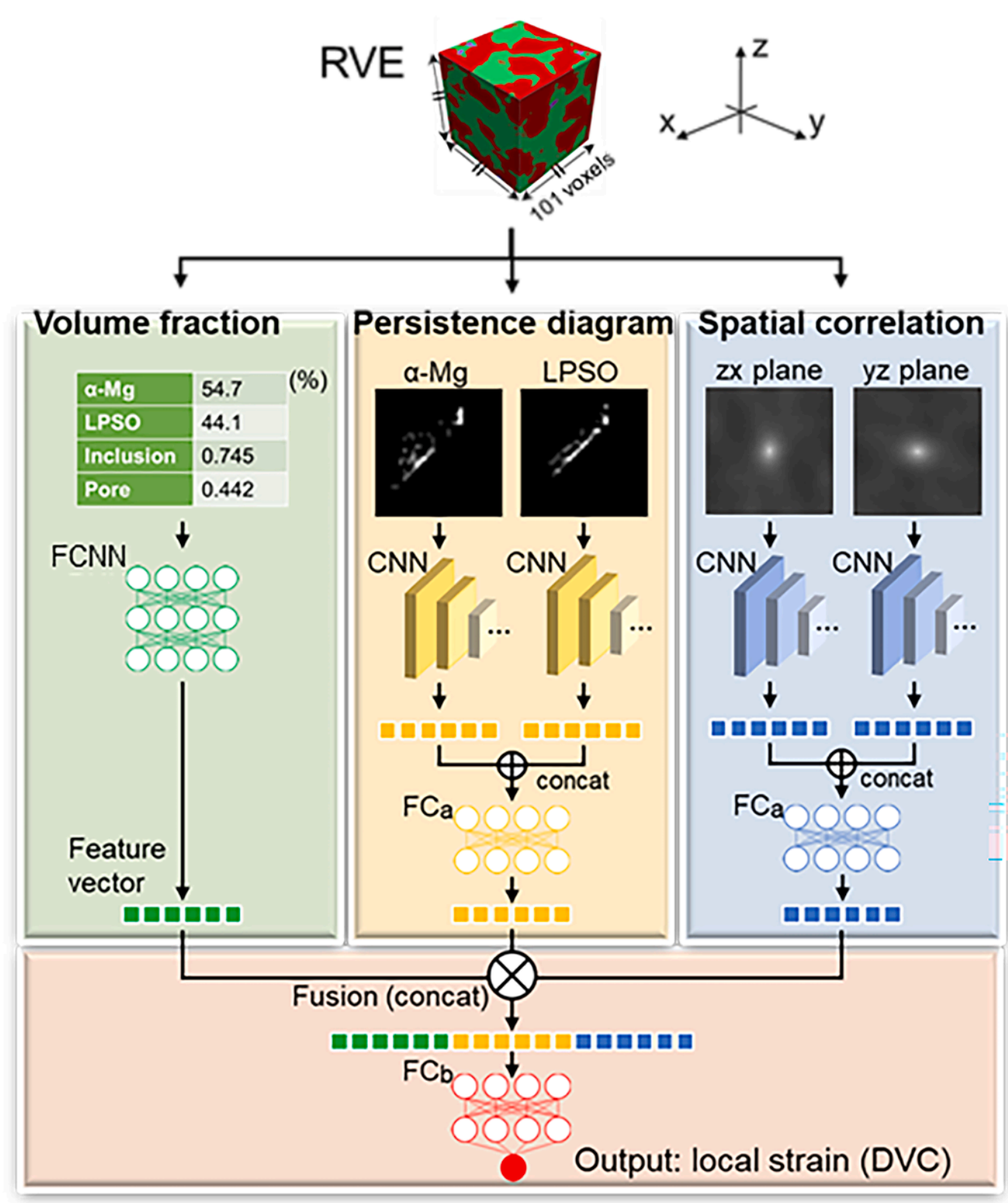

**Fig. 3.** Overall structure of multimodal deep learning model.

vector of size 192, which was then processed through fully connected layers to predict the local strain. The network structure details for each component are shown in Y

Y Table S1.

Data were divided into training and test sets in an 8:2 ratio, with 2074 training and 518 test data points. Mean Squared Error (MSE) was used as the loss function, and the Adam optimizer was employed with a learning rate of 0.001. Training was conducted over 100 epochs, ensuring convergence. The network parameters that minimized the loss on the test set were used for subsequent analyses. The model was implemented in PyTorch and trained on a GPU using CUDA.

To validate the effectiveness of the multimodal approach, separate models were also developed for each type of microstructure descriptor (volume fraction, persistence diagram, and two-point spatial correlation). Each standalone model used the feature extraction layers from the multimodal model, followed by FC layers (64→64→1) to predict local strain.

It is well-known that deep learning models can be sensitive to the initial values of network weights and biases. During initial trials, when parameters were randomly initialized, the model predominantly relied on a single microstructure descriptor, ignoring others, indicating a suboptimal solution. To mitigate this, the initial parameters for feature extraction layers were fine-tuned using the parameters from models trained on individual descriptors. This approach aimed to ensure that each descriptor contributed effectively to the final prediction.

### *2.5. Feature importance analysis*

Correlation analysis and occlusion sensitivity analysis were used to evaluate the feature importance of the microstructure descriptors and examine the relationship between the microstructure and the local strain. The following is an explanation of the two evaluation methods used. Correlation analysis is the most classical feature importance analysis. For volume fractions, which were numerical data, a scatter plot of volume fractions and local strain for each phase in the RVE was created, and correlation coefficients were calculated. For persistence diagrams and two-point spatial correlations, which were image data, the correlation coefficient between the pixel value of each pixel and the local strain was calculated and the correlation coefficient was projected to the corresponding original pixel location for all pixels.

Correlation analysis is a method for evaluating the relationship between inputs and outputs regardless of the deep learning model. As a method for visualizing the basis for model predictions, this study performed occlusion sensitivity analysis on image data. This method visualizes which parts of an image are important for prediction by observing the increase or decrease in output when a part of the image input to the model is hidden, i.e., the pixel value is set to 0. In this study, we performed an occlusion sensitivity analysis on test data using a Python library called Captum [21]. The parameters used in the calculation were the size of the window that hides the image (3×3 pixels) and the amount of window movement (2 pixels). The results of the Occlusion Sensitivity Analysis were obtained for each of the 512 test data, and the results were averaged over all of them.

## 3. Results and discussion

### 3.1. Strain distribution obtained by DVC

Fig. 4 shows the X-ray CT results of as-cast $Mg_{94}Zn_2Y_4$. In the X-ray CT images, the phases appear whiter when the X-ray absorptivity of the phases is high, i.e., when the phases are dense or contain heavy elements. Thus, the white phase in Fig. 4 corresponds to Y-rich inclusions, the light gray phase to the LPSO phase which is $Mg_{85}Zn_6Y_9$ [22], the dark gray phase to the α-Mg phase which is $Mg_{99.2}Zn_{0.2}Y_{0.6}$ [22], and the black phase to pores that are air.

Fig. 5 shows the 3D equivalent strain distribution obtained by DVC when compressed to 300 MPa. The red area indicates the area where high strain occurred, and it was clear that macroscopic banded high strain regions occurred. In addition, it was found that a macrostrain of about 5% occurred when the specimen was compressed by 300 MPa, but in the high-strain region, the strain was about 3-4 times larger than the macrostrain. The displacement distribution obtained from DVC and the distribution of each strain component are shown in the Supplementary Figures S2 and S3.

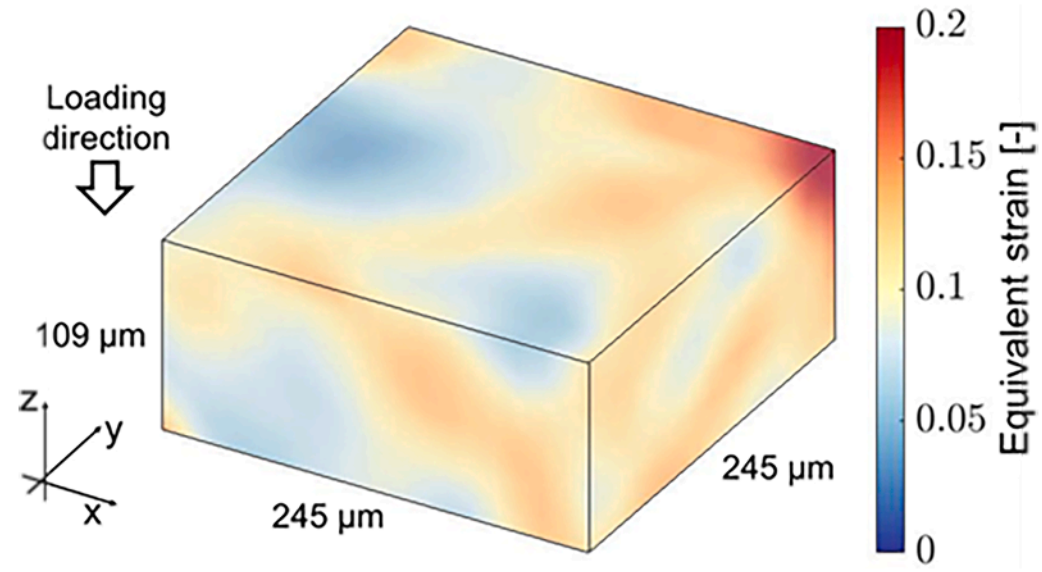

**Fig. 5.** 3D equivalent strain field obtained by DVC (300 MPa compressive stress and 5% macroscopic strain).

### 3.2. Prediction results by deep learning

The accuracy results of predicting local strain from microstructure descriptors on test data are shown in Fig. 6. Among the single microstructure descriptors, the most accurate predictors, in descending order, were the two-point spatial correlation, persistence diagram, and volume fraction. The superior performance of the two-point spatial correlation is attributed to its ability to represent the spatial distribution of phases and consider the direction of elongation, which the persistence diagram cannot evaluate. Furthermore, when comparing predictions made using individual descriptors to those made using a multimodal model that incorporated all three descriptors, the multimodal model demonstrated significantly higher prediction accuracy than the single descriptor model that utilized the two-point spatial correlation. This enhanced accuracy is attributed to the multimodal model's ability to combine microstructural features from the two-point spatial correlation with macrostructural features from the volume fractions and detailed phase connectivity from the persistence diagrams. This integration of diverse information sources facilitated a more comprehensive prediction.

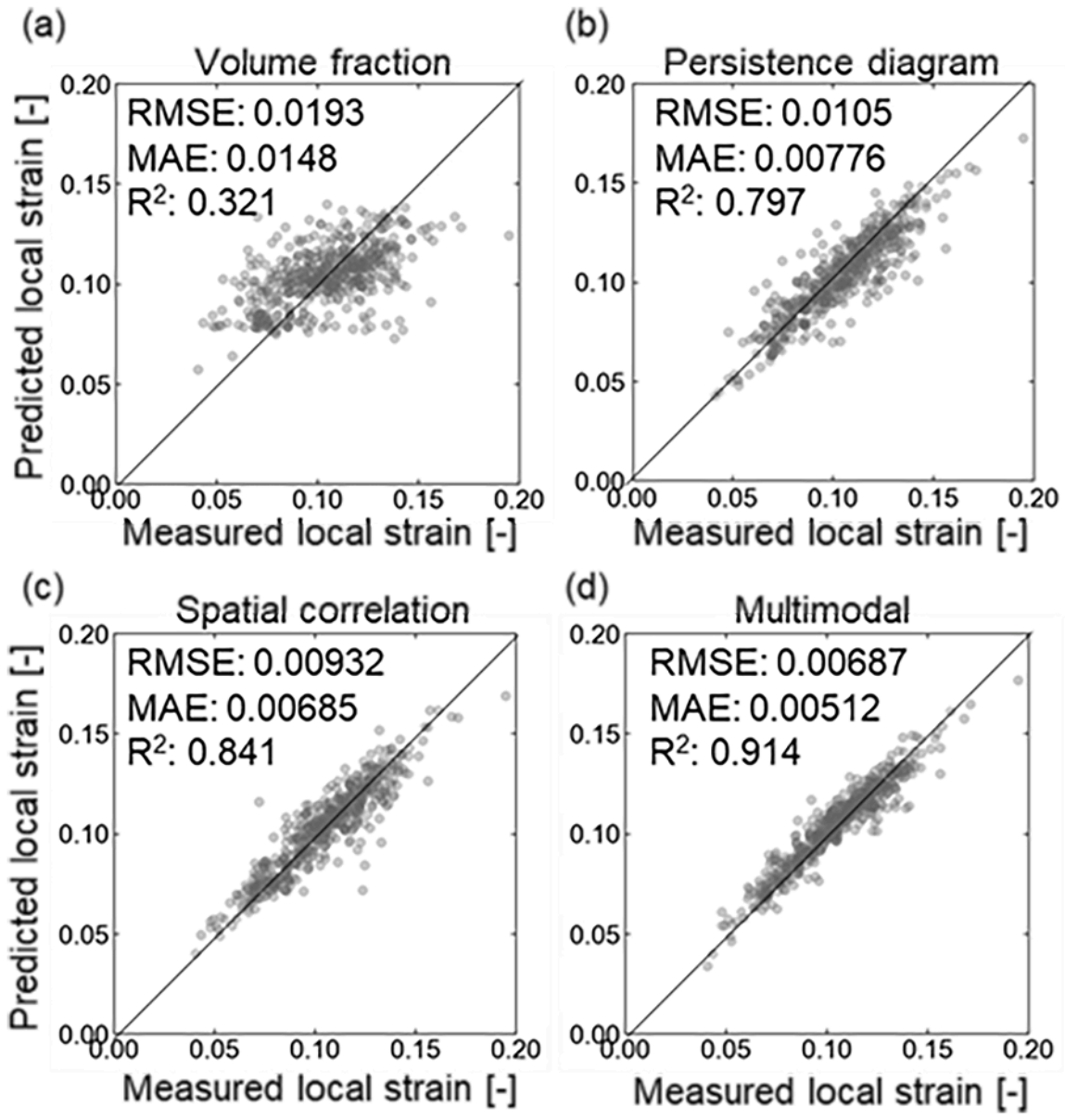

**Fig. 6.** Comparison between measured local strain by DVC and predicted local strain by deep learning model. Accuracy indices are shown in the upper left corner of each figure, where RMSE stands for root mean squared error, MAE for mean absolute error, and R2 for coefficient of determination. (a) Prediction result by volume fraction only, (b) Prediction result by persistence diagram only, (c) Prediction result by two-point spatial correlation only, and (d) Prediction result by multimodal model.

To further validate the predictive performance of the multimodal model, the predicted equivalent strain distributions in the XY, XZ, and YZ cross-sections were plotted alongside the equivalent strain measured by DVC in Fig. 7. The plots demonstrate that the model accurately captures strain localization behaviors, such as the concentration of strain in the upper right area and the limited deformation in the central left area of the XY cross-section. These results suggest that, for this material, strain localization can be predicted with a reasonable degree of accuracy based on the two-phase distribution information in the microstructure.

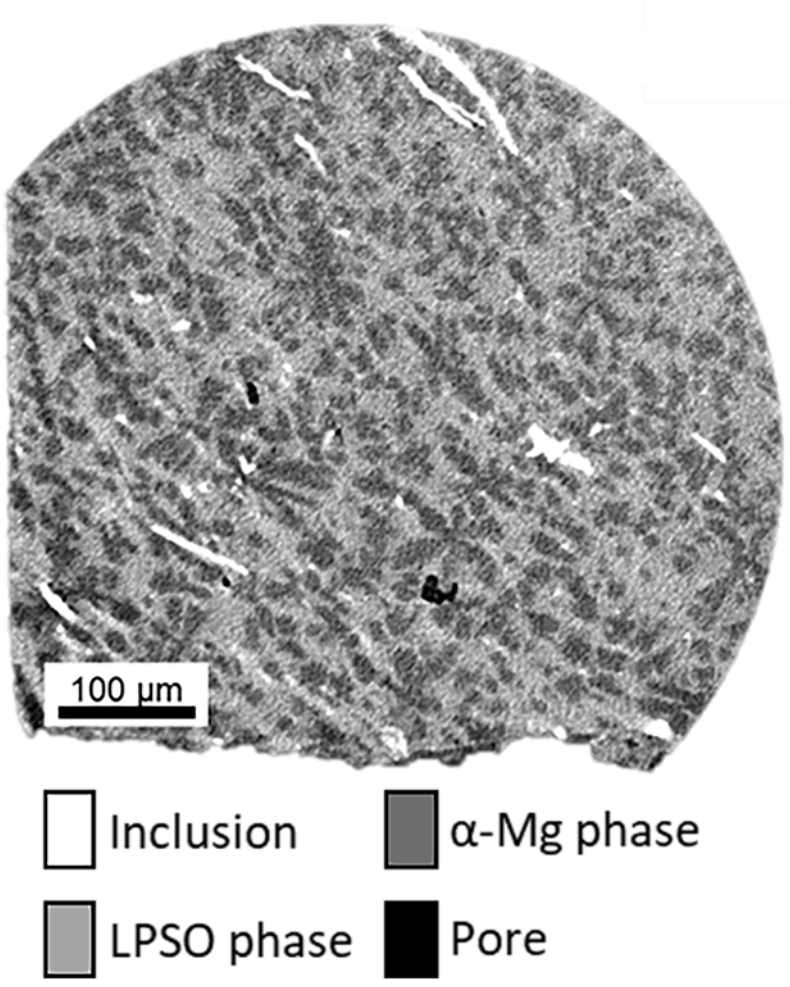

**Fig. 4.** CT-sliced image of Mg/LPSO two-phase alloy.

### 3.3. Feature importance analysis

#### 3.3.1. Effect of volume fraction

The results of the correlation analysis are shown in Fig. 8, which shows a scatter plot of the volume fraction of each phase and local strain in RVE. The volume fraction of α-Mg phase and local strain had a weak negative correlation (Fig. 8 (a)), while the volume fraction of LPSO phase and local strain had a weak positive correlation (Fig. 8 (b)). This indicates that the hard LPSO phase is preferentially plastically deformed in this material, which is consistent with the HR-DIC results from a previous study [3]. In addition, a trend was observed in Fig. 8 (c) where high strain values were more common when the volume fraction of inclusions in the RVE was less than 2.5%. When the volume fraction of inclusions exceeded 2.5%, the local strain values were more consistently lower, ranging between 0.05 and 0.1. This behavior is attributed to the Y-rich inclusions [23], which are harder and less deformable than the matrix phase, thereby reducing the local strain. Regarding the volume fraction of pores, there was little correlation with local strain (Fig. 8 (d)). In a previous study that applied DVC to die-cast aluminum alloys, pores had a strong influence on strain localization [24]. However, in the material studied in this research, the pores were very small (10-20 μm in diameter at the largest), and their quantity was limited, leading to a limited effect of porosity on strain localization. In Fig. 8 (d), two groups centered around local strain values of approximately 0.12 and 0.08 were observed. The former likely corresponds to the high-strain band observed in Fig. 5, while the latter corresponds to the remaining regions.

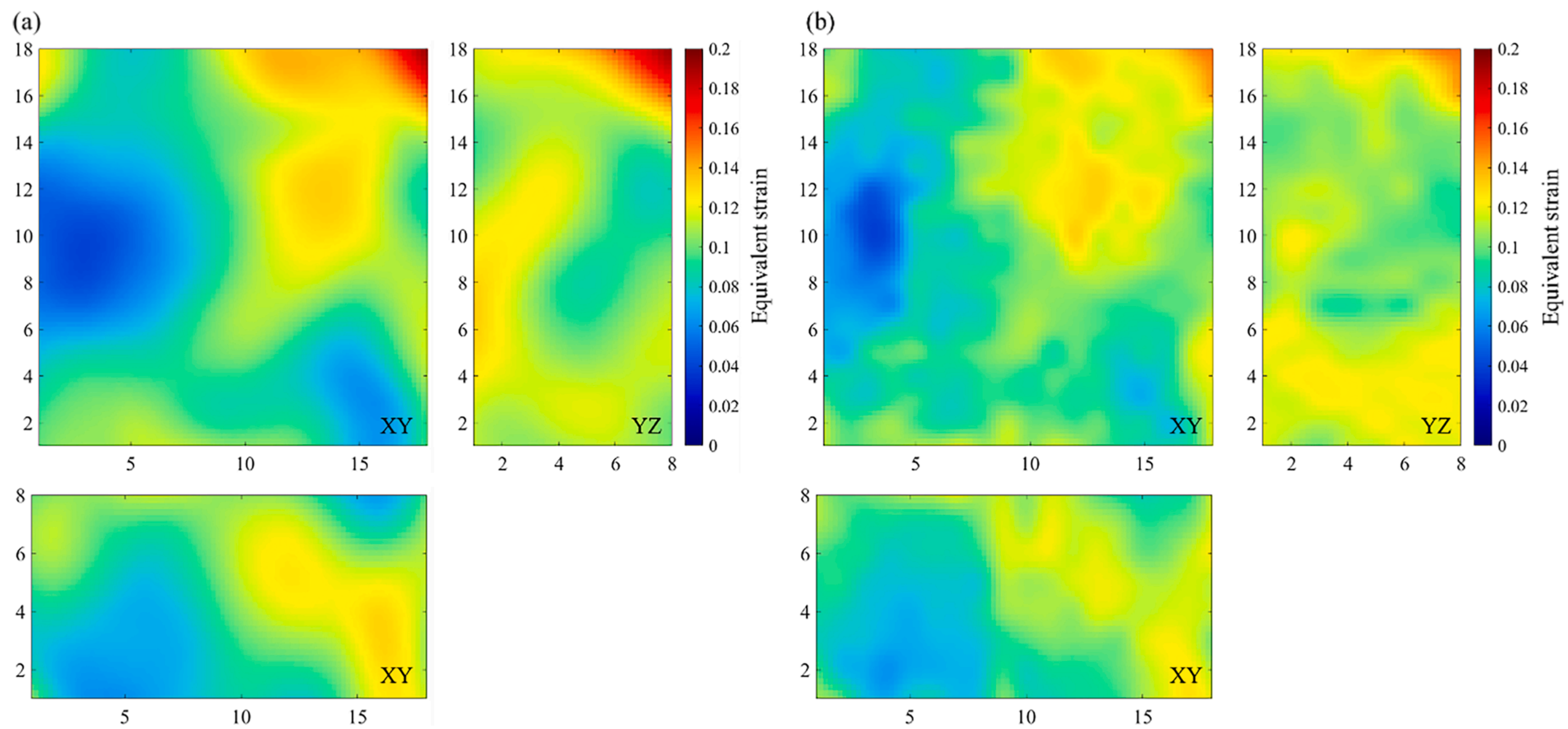

**Fig. 7.** Equivalent strain distribution obtained from (a) DVC and (b) multimodal deep learning, shown in the XY, XZ, and YZ sections. The loading direction is the Z direction. The units for each coordinate are in voxels (1 voxel = 0.68 μm).

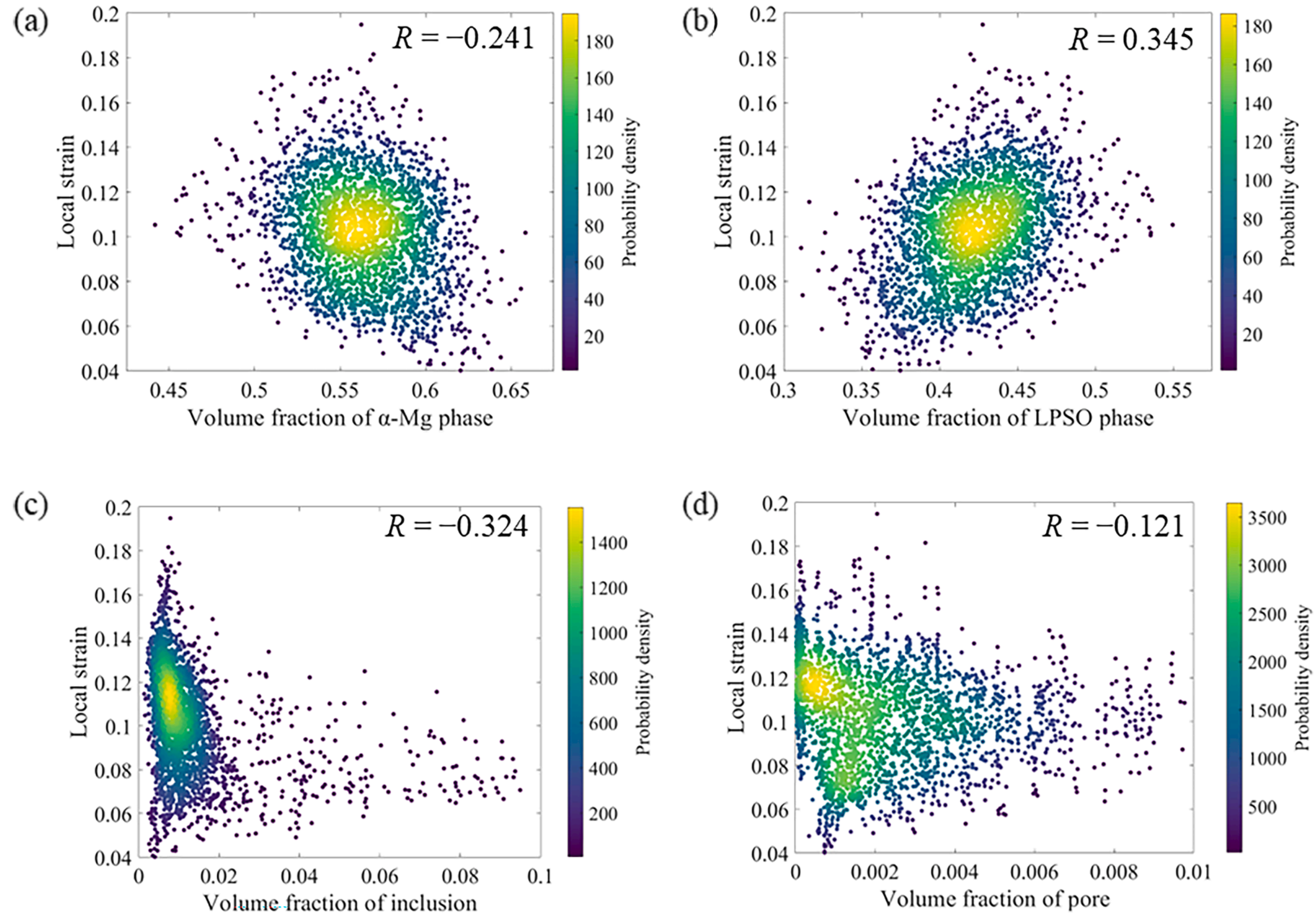

**Fig. 8.** Scatter plots of volume fraction of each phase in the RVE and local strain. The numbers in the upper right of each figure represent correlation coefficients. (a) α-Mg phase, (b) LPSO phase, (c) inclusion and (d) pore.

The correlation between microstructure and mechanical response as discussed above generally depends on RVE size. In this study, the RVE was set as a cube with 101 voxels per side. Here, we investigated the change in the relationship between the volume fraction of the LPSO phase and the local strain when the size of the RVE was varied: the length of one side of the RVE was varied by 5 voxels from 21 to 241 voxels, and the correlation coefficient between the volume fraction of the LPSO phase in the RVE and the local strain was calculated. Fig. 9 shows the relationship between the length of one side of the RVE and the correlation coefficient. The correlation coefficient was found to increase linearly from a value almost equal to zero, and remained around 0.35-0.38 after the length of one side of the RVE became 101 voxels, which was used in this study. There is a trade-off relationship between spatial resolution and accuracy of correlation analysis, making it crucial to determine the RVE size based on the microstructural scale.

### 3.3.2. *Effect of phase connectivity*

Phase connectivity is quantified using the death values from the 0th persistence homology analysis. Fig. 10 shows the persistent diagrams for both the α-Mg phase and the LPSO phase, calculated from the overall phase distribution in the observed region, along with the corresponding histograms of the death values. In Fig. 10 (b) and (c), the horizontal axis "Birth" indicates the size of the phase, with a smaller value corresponding to a larger phase, while the vertical axis "Death" indicates the connectivity of the phase, with a smaller value corresponding to a more connected phase (See Fig. 3 of the paper in [25]). In Fig. 10 (d), the histogram is plotted for data with a lifetime (= death − birth) greater than or equal to 2 to reduce noise effects. The average death values for the α-Mg phase and the LPSO phase are −8.17 and −5.75, respectively, indicating that the α-Mg phase, with a larger volume fraction, has higher connectivity.

The results of the correlation analysis of the persistence diagrams are shown in Fig. 11. Red points indicate a positive correlation with local strain, while blue points indicate a negative correlation with local strain. The regions in the α-Mg phase where both Birth and Death values are high (circled in red in Fig. 11 (a)) and the regions in the LPSO phase where both Birth and Death values are low (circled in red in Fig. 11 (b)) each show a positive correlation with local strain. In other words, small island-shaped α-Mg phases and large, densely connected LPSO phases were found to be associated with high strain. Additionally, in corresponding regions of the other phase, the positive and negative correlations are reversed, indicating that the connectivity of the phases has opposite effects on local strain.

In other Mg alloys besides the Mg/LPSO two-phase alloys, differences in strain localization behavior and associated elongation due to phase connectivity were discussed in Mg-Li-Gd dual-phase alloys consisting of α-Mg and β-Li phases [26]. In Mg-5Al-3Ca alloys, strength and ductility also varied depending on the connectivity of the intermetallic Laves phase in the material [27]. Thus, the connectivity of phases affects the properties and mechanical behavior of materials. Although many studies are currently limited to the qualitative observation of phase connectivity, quantitative evaluation of connectivity will become important in the future, and persistent homology analysis is expected to be utilized in this regard.

In this study, the correlation coefficients between each point (each pixel) on the persistence diagram and the objective variable were obtained, and they were projected on the same scale as the persistence diagram. In a similar method, Obayashi et al. [25] combined persistence diagram and linear regression, and proposed a method to visualize important points on the persistence diagram by reconstructing the learned weights of the linear regression on the persistence diagram. This method has also been used in materials science. For example, Kimura et al. [11] used this method to identify microstructure features that are likely to be crack initiation sites in iron ore sinters during reduction. However, this method assumes linear regression, which is difficult to apply to complex tasks. In such cases, the method proposed in this study is useful because it can easily reveal points on the persistent diagram that are strongly related to the objective variable.

### 3.3.3. *Effect of phase orientation*

The results of the correlation analysis of the two-point spatial correlation are shown in Fig. 12. In both the zx plane and yz plane, the regions oblique to the direction of loading had a high positive correlation with strain, especially in the zx plane, the region 45° to loading direction. In Mg/LPSO two-phase alloys, it is known that the (0001) basal plane of the LPSO phase tends to be oriented parallel to the elongation direction, and when the (0001) basal plane is inclined at 45 ° to the loading direction, the Schmid factor of the (0001) $<11\bar{2}0>$ primary slip system becomes maximal, making basal slip more likely to occur [2]. In addition, as shown in Fig. 1 (a), the size of the microstructure used for DVC in this study was such that many grains existed. Within these grains, there were microstructures elongated at 45° to the loading direction, and significant deformation was expected to occur in such structures. Considering these, the results in Fig. 12 are deemed reasonable. In summary, the results of the correlation analyses indicated that high strain tended to occur in the region where LPSO phases had a large elongated phase in the 45° direction of the loading direction. In research utilizing HR-DIC, high strain was observed in elongated LPSO grains [3], which enhances the validity of the results obtained in this study.

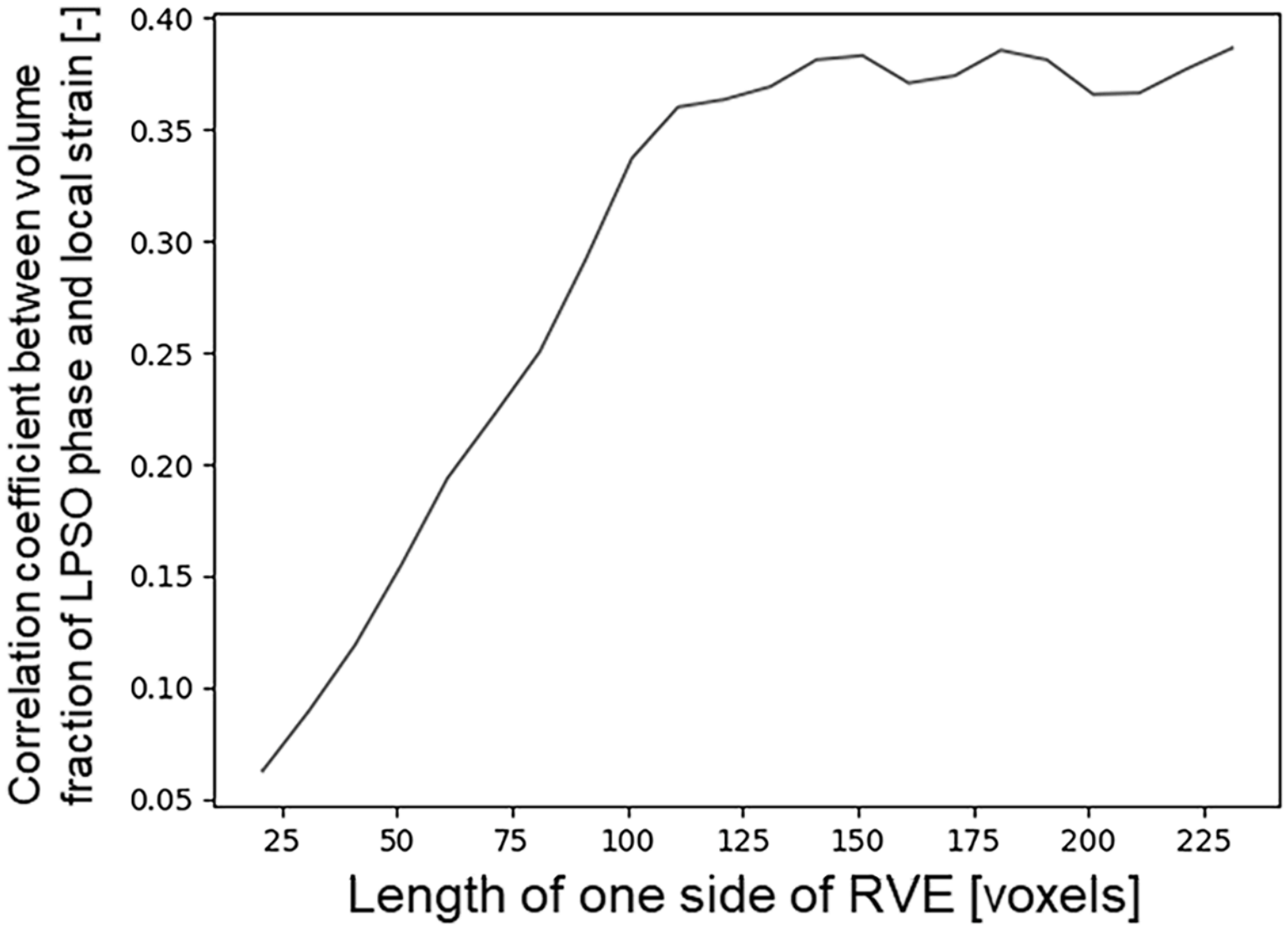

**Fig. 9.** Relationship between length of one side of RVE and correlation coefficient between volume fraction of LPSO phase and local strain (1 voxel was about 0.68 μm).

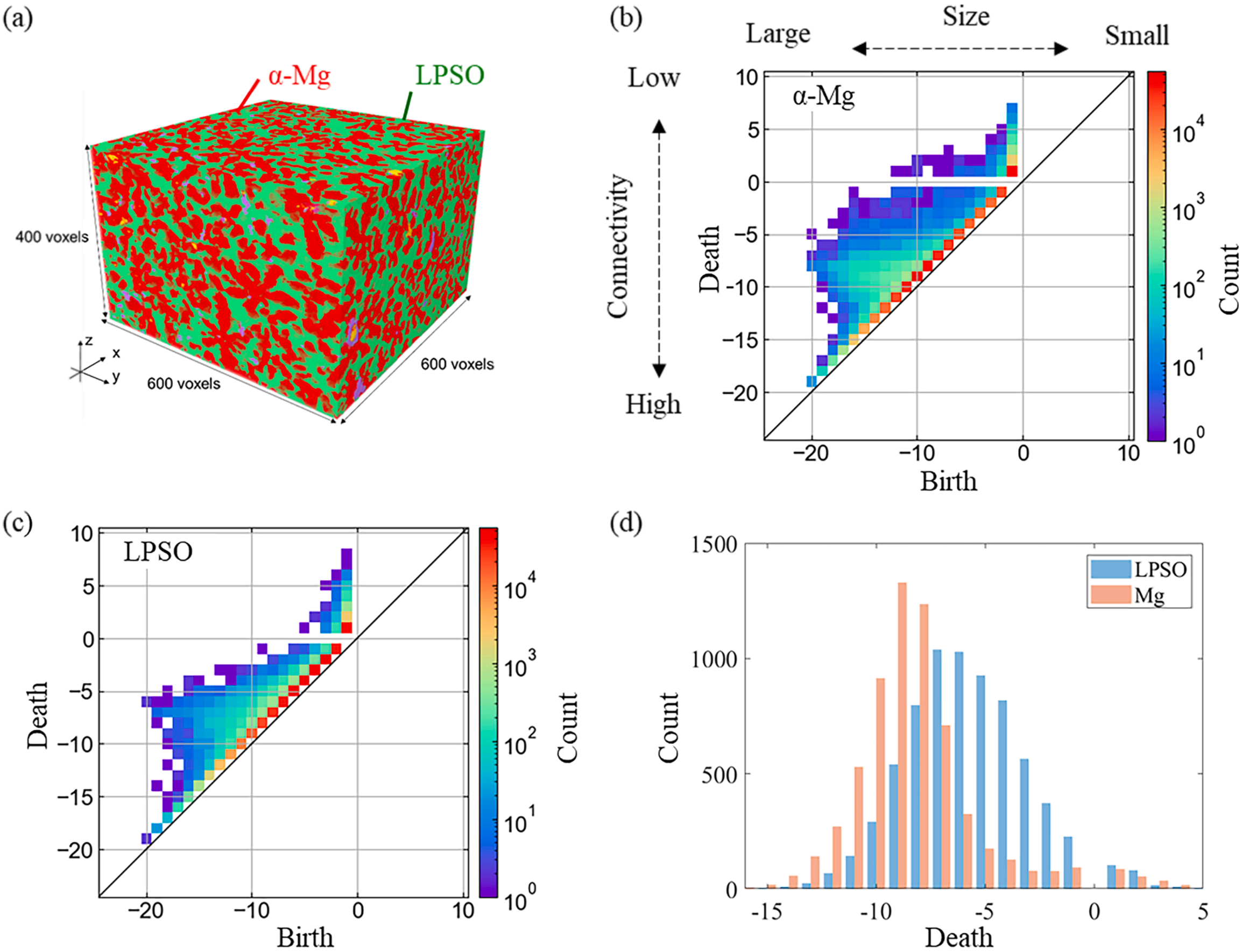

**Fig. 10.** (a) Overall phase distribution in the observed region, (b) Persistent diagram of the α-Mg phase, (c) Persistent diagram of the LPSO phase, (d) Histogram of death values for data with a lifetime (= birth − death) greater than or equal to 2.

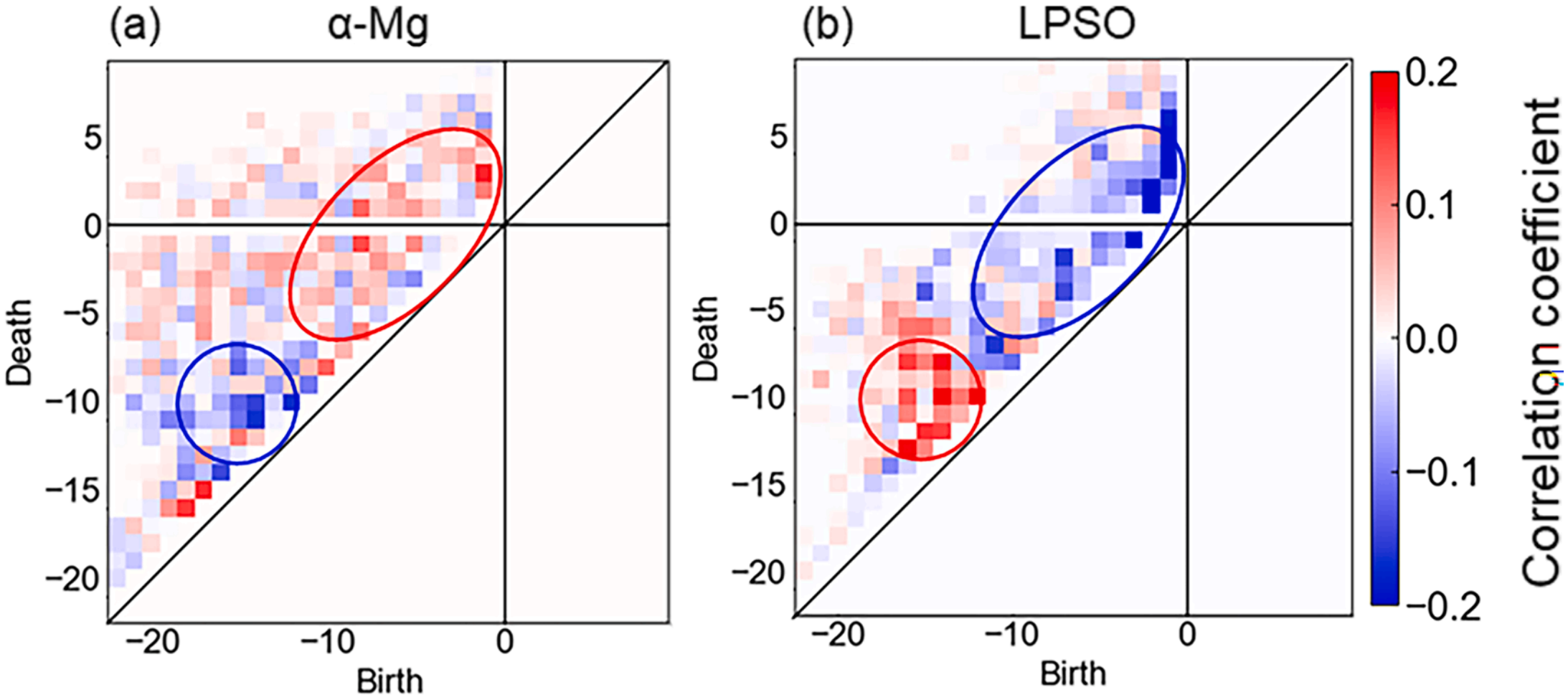

**Fig. 11.** Results of correlation analysis in persistence diagram. Red dots are positively correlated with local strain and blue dots are negatively correlated with local strain: (a) α-Mg phase and (b) LPSO phase.

Additionally, the high correlation coefficient observed in Fig. 12 for only one of the 45° axes in the loading direction may be attributed to the macroscopic non-uniform deformation of the specimen or the influence of surface constraints due to measuring the strain distribution in the region near the specimen surface.

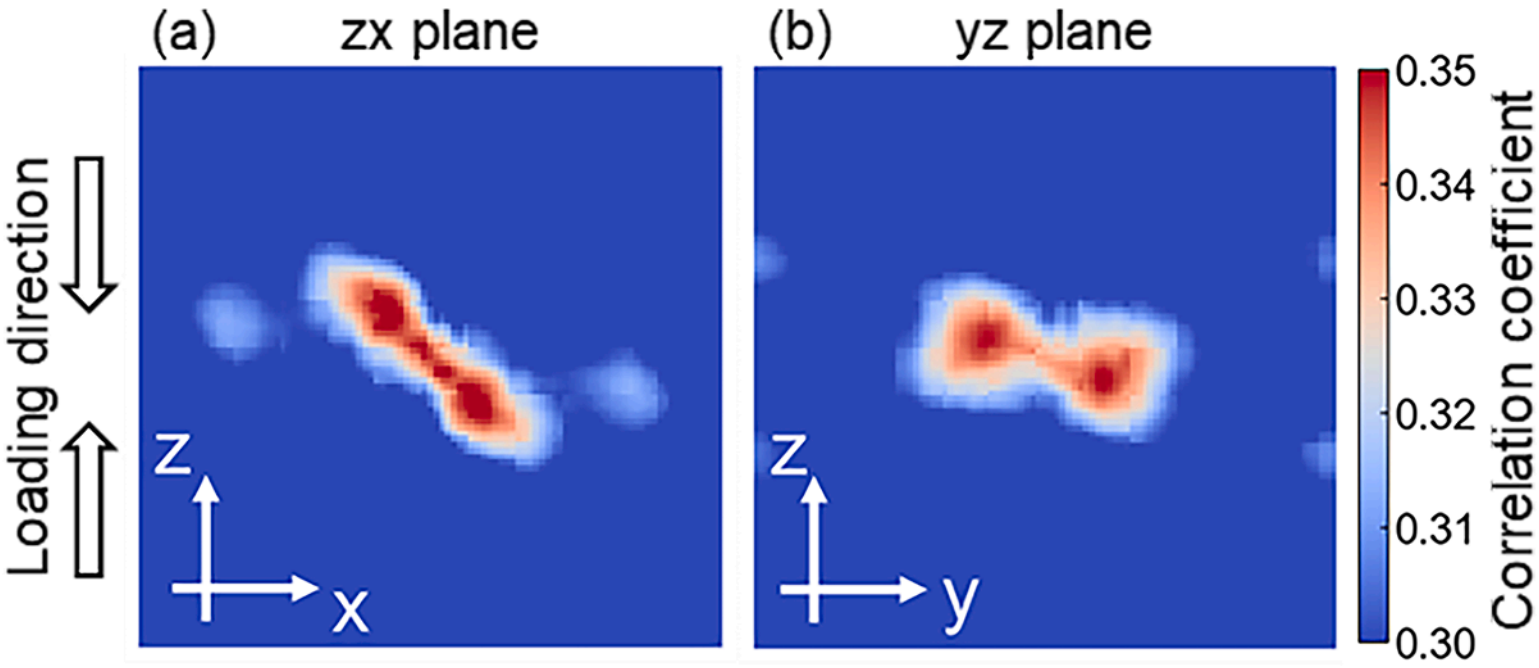

**Fig. 12.** Results of correlation analysis in two-point spatial correlation. Red areas have higher positive correlation with local strain. (a) zx plane and (b) yz plane.

Fig. 13 shows the results of the occlusion sensitivity analysis performed to evaluate the feature importance. In the persistence diagram of the LPSO phase, regions where both Birth and Death are small had a positive effect on strain, while regions where both Birth and Death are large had a negative effect on strain. In the two-point spatial correlation, the red region near the center of the image had a similar shape to the correlation analysis results. In other words, the results of the occlusion sensitivity analysis were similar to those of the importance evaluation by the correlation analysis for both the persistence diagram and the two-point spatial correlation. However, in the occlusion sensitivity analysis, there were areas where positive and negative influences were mixed in a complex manner, indicating that the complex learning resulted in highly accurate predictions. Comparing Fig. 13 (a) and (b), it was also found that the increase and decrease in output were larger in the persistence diagram of the LPSO phase, indicating that the connectivity of the LPSO phase was more important than that of the α-Mg phase for prediction.

The proposed approach enables the quantification of phase connectivity in multiphase microstructures and its correlation with local strain. This analysis is particularly relevant for advanced structural materials, such as advanced steels containing multiple phases, dual-phase titanium alloys, high-entropy alloys, metal-metal nanocomposites, and ceramic matrix composites. These materials often contain multiple phases, and their mechanical reliability is strongly influenced by the spatial distribution of these phases. By quantifying phase connectivity and analyzing its relationship with local strain, the proposed approach can provide valuable insights into the mechanisms of deformation and failure. This approach can be easily extended and appled to a wide range of studies aiming to predict material properties and mechanical behavior from various information about materials. For example, focused ion beam machining (FIB)-SEM provides a method for acquiring 3D microstructure images with higher spatial resolution than X-ray CT, and incorporating 3D EBSD allows obtaining 3D crystallographic information [28, 29]. Additionally, the crystal plasticity finite element method (CPFEM) has been applied to various materials as a method to predict material properties and mechanical behavior from 3D microstructure data [30-32]. The multimodal deep learning approach introduced in this study can serve as a means to establish linkages between 3D microstructure data (microstructure morphology and crystallographic information) and material properties using CPFEM.

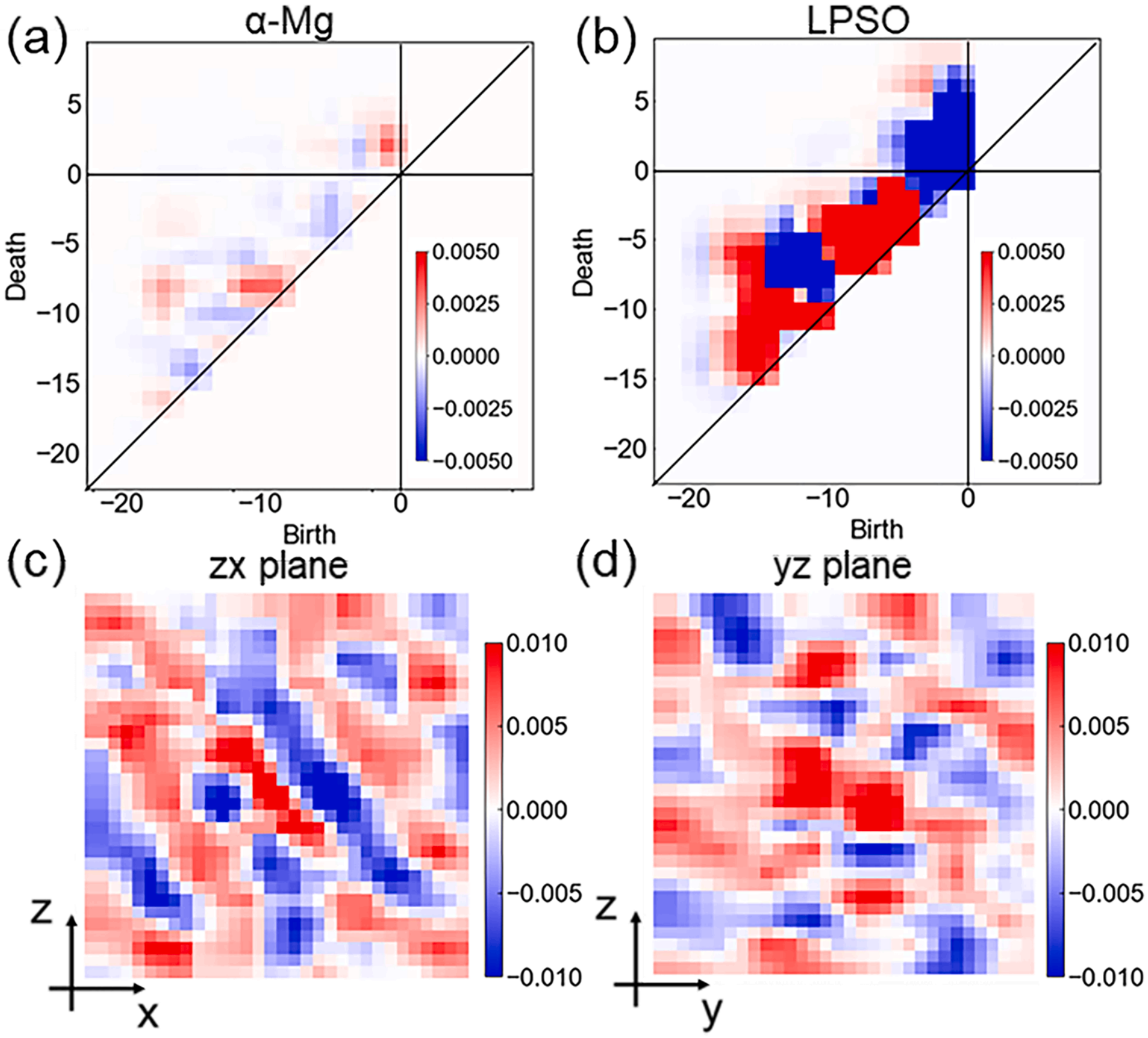

**Fig. 13.** Results of occlusion sensitivity analysis. The color map value is the difference between the original output value and the output value when a certain area is hidden: (a) persistence diagram of α-Mg phase, (b) persistence diagram of LPSO phase, (c) zx plane and (d) yz plane.

## 4. Conclusions

In this study, DVC was applied to X-ray CT images of Mg/LPSO two-phase alloys to measure 3D strain distributions. We also proposed a method to reveal the relationship between 3D microstructure and local strain. The following conclusions can be drawn:

(1) In the Mg/LPSO two-phase alloy, the density differences between the constituent phases were sufficient to identify each phase on X-ray CT. The DVC was successfully applied to the X-ray CT images because sufficient patterns existed in the images, and the DVC enabled measurement of the 3D strain distribution under compressive deformation, indicating that macroscopic banded high strain regions occurred.
(2) Three microstructure descriptors, volume fraction of each phase, persistence diagram, and two-point spatial correlation, were extracted from 3D microstructure images around the strain measurement points. The persistent diagram quantifies phase connectivity, while the two-point spatial correlation quantifies phase orientation. A multimodal deep learning model was constructed to predict local strain from the extracted microstructure descriptors with different formats. By applying fine-tuning to simplify learning, the multimodal deep learning model showed higher prediction accuracy than when predicting from only a single microstructure descriptor.
(3) To elucidate the relationship between microstructure and local strain, correlation analysis and occlusion sensitivity analysis, a method for visualizing the predictive basis of deep learning, were conducted. Both methods showed that high strain tended to occur in the region where LPSO phases had large elongated phases in the 45° direction of the loading direction in as-cast Mg/LPSO two-phase alloys $Mg_{94}Zn_2Y_4$, a result consistent with previous studies, indicating the effectiveness of the proposed method in this study.

## Data availability statement

The X-ray CT images, image segmentation results, DVC results, and Python codes used for performing the multimodal deep learning are available at the following link: https://gitlab.rme.mm.t.u-tokyo.ac.jp/pub/multimodalai.

## CRediT authorship contribution statement

**Daiki Kuriki:** Writing – original draft, Visualization, Validation, Software, Methodology, Formal analysis, Conceptualization. **Fabien Briffod:** Writing – review & editing, Software, Formal analysis. **Takayuki Shiraiwa:** Writing – review & editing, Supervision, Methodology, Investigation, Funding acquisition, Conceptualization. **Manabu Enoki:** Writing – review & editing, Supervision, Funding acquisition, Conceptualization.

## Declaration of competing interest

The authors declare that they have no known competing financial interests or personal relationships that could have appeared to influence the work reported in this paper.

## Acknowledgments

This work was supported by JSPS KAKENHI Grant Number 23H04464 and the joint usage/research grant of Institute of Light Metals (ILM) at Kumamoto University and University of Toyama.

## Supplementary materials

Supplementary material associated with this article can be found, in the online version, at doi:10.1016/j.actamat.2024.120398.